\documentstyle[12pt]{article}

\begin{document}

\begin{titlepage}

\title{New variants of the Bell-Kochen-Specker theorem\footnote{Phys. Lett. A 
{\bf 218}, 115 (1996).}}

\author{Ad\'{a}n Cabello\thanks{Electronic address: fite1z1@sis.ucm.es}\\
Jos\'{e} M. Estebaranz\\
Guillermo Garc\'{\i}a-Alcaine\\
{\em Departamento de F\'{\i}sica Te\'{o}rica,}\\
{\em Universidad Complutense, 28040 Madrid, Spain.}}

\date{May 20, 1996}

\maketitle

\begin{abstract}
We discuss two new demonstrations of the Bell-Kochen-Specker theorem: a
state-independent proof using 14 propositions in ${\bf R}^4$, based on a
suggestion made by Clifton, and a state-specific proof involving 5
propositions on the singlet state of two spin-$\frac{1}{2}$ particles.

\end{abstract}

\end{titlepage} 

We recently found \cite{1} a proof of the Bell-Kochen-Specker (BKS in the
following) theorem \cite{2,3} using 18 vectors in ${\bf R}^4$. While the
paper was in press, Clifton \cite{4} suggested to us a modification allowing
a further reduction of this number. In the first part of this paper we will
discuss Clifton's idea. In the second part we will take advantage of the
properties of the singlet state of two spin-$\frac{1}{2}$ particles to find a
state-specific BKS\ proof involving 5 propositions.

The BKS theorem asserts the impossibility of hidden variables theories such
that the values $v\left( {\bf u}_i\right) $ of propositions $P_{{\bf u}_i}$
represented by projectors $\left| {\bf u}_i\right\rangle \left\langle {\bf u}%
_i\right| $ obey the following requisites:
\begin{enumerate}
\item[(a)]  {\it In an individual system each proposition $P_{{\bf u}_i}$
has a unique value, }$0${\it \ (``no'') or }$1$ {\it (``yes''),\ that is
independent of any other compatible observables being considered jointly
(non-contextuality). }
\item[(b)]  {\it For each set of rank 1 projectors whose sum is the unit
matrix (in the }${\it n}${\it -dimensional Hilbert space of the states of
the system), the value of one and only one of the corresponding propositions
is }$1${\it , and the values of the remaining }$n-1${\it \ propositions are }%
$0${\it .}
\end{enumerate}
Conditions (a) and (b) imply a third one:
\begin{enumerate}
\item[(c)]  {\it Given two sets of propositions represented by projectors
over the same subspace, }$\sum\limits_i\left| {\bf u}_i\right\rangle
\left\langle {\bf u}_i\right| =\sum\limits_i\left| {\bf v}_i\right\rangle
\left\langle {\bf v}_i\right| ${\it , the sums of values must be equal, $%
\sum\limits_iv\left( {\bf u}_i\right) =\sum\limits_iv\left( {\bf v}_i\right) 
$. }
\end{enumerate}
The implication is evident: if we have two different sets of projectors
summing up to the unit matrix, $\sum\limits_{i=1}^n\left| {\bf u}%
_i\right\rangle \left\langle {\bf u}_i\right| =\sum\limits_{i=1}^n\left| 
{\bf v}_i\right\rangle \left\langle {\bf v}_i\right| ={\bf 1}$, and both
sets share a common element, say ${\bf u}_1={\bf v}_1$, premise (a) implies
that the value of the corresponding proposition in both sets must be the
same ($0$ or $1$); then premise (b) implies that the sums of the values of
both sets of propositions on the complementary subspace must also be equal ($%
1$ or $0$, respectively), $\sum\limits_{i=2}^nv\left( {\bf u}_i\right)
=\sum\limits_{i=2}^nv\left( {\bf v}_i\right) $. The same argument can be
used again if more vectors are common.

By choosing among eqs. (1--9) of ref. \cite{1} four couples of equations,
each couple with a common vector, and equating the sums of values over the
subspaces complementary to these common vectors, we reduce the system to
only 5 equations with 14 vectors, for instance, 
\begin{equation}
\label{e1}
\begin{array}{c}
v\left( 0,0,1,0\right) +v\left( 1,1,0,0\right) +v\left( 1,-1,0,0\right) = \\ 
v\left( 0,1,0,0\right) +v\left( 1,0,1,0\right) +v\left( 1,0,-1,0\right) , 
\end{array}
\end{equation}
\begin{equation}
\label{e2}
\begin{array}{c}
v\left( 1,-1,-1,1\right) +v\left( 1,1,0,0\right) +v\left( 0,0,1,1\right) =
\\ 
v\left( 1,1,1,1\right) +v\left( 1,0,-1,0\right) +v\left( 0,1,0,-1\right) , 
\end{array}
\end{equation}
\begin{equation}
\label{e3}
\begin{array}{c}
v\left( 0,0,1,0\right) +v\left( 0,1,0,0\right) +v\left( 1,0,0,1\right) = \\ 
v\left( 1,-1,-1,1\right) +v\left( 1,1,1,1\right) +v\left( 0,1,-1,0\right) , 
\end{array}
\end{equation}
\begin{equation}
\label{e4}
\begin{array}{c}
v\left( 1,1,-1,1\right) +v\left( 1,0,1,0\right) +v\left( 0,1,0,-1\right) =
\\ 
v\left( 1,1,1,-1\right) +v\left( 1,0,0,1\right) +v\left( 0,1,-1,0\right) , 
\end{array}
\end{equation}
\begin{equation}
\label{e5}v\left( 1,1,-1,1\right) +v\left( 1,1,1,-1\right) +v\left(
1,-1,0,0\right) +v\left( 0,0,1,1\right) =1. 
\end{equation}

We can now formulate the following version of the BKS\ theorem:
\begin{quotation}
{\it There is no set of values }$v\left( {\bf u}_i\right) ${\it \ verifying
eqs. (\ref{e1}--\ref{e5}).}
\end{quotation}
The proof involves a parity argument: if we add these five equations, each
value $v\left( {\bf u}_i\right) $ appears either twice on the same side of
the resulting equation, with an even contribution ($0$ or $2$), or once on
each side, with a cancellation of both contributions; the extra term 1 on
the right-hand side makes it impossible to satisfy the equality.

This 14-vector set (or any of the many others that we can obtain similarly)
leads to a proof of the BKS theorem based on the explicit use of (c),
according to Clifton's suggestion. Condition (c) is a direct consequence of
(a) and (b), and does not impose any new requirement on hidden variables.
Its use is not an artifice to leave out some propositions when counting the
number of them involved in the proof, because no concrete value for the
propositions eliminated is assumed; the proof stands, whatever the values of
the omitted propositions.

Let us now consider a system of two spin-$\frac{1}{2}$ particles and choose the basis
formed by the eigenvectors of $\sigma _z^{(1)}\otimes \sigma _z^{(2)}$
(i.e., the vectors up$\otimes $up, up$\otimes $down, down$\otimes $up, down$%
\otimes $down). In any individual system prepared in the singlet state, $%
\left( 0,1,-1,0\right) $ (we omit, as before, the normalization constant),
we have $v\left( 0,1,-1,0\right) =1$, and the values for propositions over
any orthogonal direction must be zero ($v\left( 1,1,1,-1\right) =0$, $%
v\left( -1,1,1,1\right) =0$, for instance). Replacing these values into eqs.
(7,8) of ref. \cite{1} (or into eqs. (\ref{e5}) and (\ref{e4}) in this
paper, respectively, but the intermediate step to obtain (\ref{e4}) from
(8,9) in \cite{1} is actually unnecessary) we obtain 
\begin{equation}
\label{e6}v\left( 1,1,-1,1\right) +v\left( 1,-1,0,0\right) +v\left(
0,0,1,1\right) =1,
\end{equation}
\begin{equation}
\label{e7}v\left( 1,1,-1,1\right) +v\left( 1,0,1,0\right) +v\left(
0,1,0,-1\right) =1.
\end{equation}
The hidden variables values for the four non-repeated propositions in (\ref
{e6},\ref{e7}) satisfy the following relation in the singlet state: 
\begin{equation}
\label{e8}v\left( 1,-1,0,0\right) +v\left( 0,0,1,1\right) +v\left(
1,0,1,0\right) +v\left( 0,1,0,-1\right) =1.
\end{equation}
The proof of (\ref{e8}) is straightforward: first, the value of a
factorizable proposition (like the ones appearing in (\ref{e8})) is the
product of the values of its factors, $v\left( \left( a,b\right) ^{\left(
1\right) }\otimes \left( c,d\right) ^{\left( 2\right) }\right) =$ $v\left(
\left( a,b\right) ^{\left( 1\right) }\right) \,v\left( \left( c,d\right)
^{\left( 2\right) }\right) $\footnote{%
Asking if the two-particle system is in the state $\left( a,b\right)
^{\left( 1\right) }\otimes \left( c,d\right) ^{\left( 2\right) }$ is the
same as asking if particle 1 is in the state $\left( a,b\right) $ {\it and}
particle 2 is in the state $\left( c,d\right) $: the answer is 1 (yes) only
if the values of both factor propositions are 1, and is 0 otherwise. The
same conclusion can be reached as a consequence of the Kochen-Specker
product rule for compatible observables $v\left( AB\right) =v\left( A\right)
v\left( B\right) $; in particular $v\left( \left| u\right\rangle
\left\langle u\right| ^{\left( 1\right) }\otimes \left| w\right\rangle
\left\langle w\right| ^{\left( 2\right) }\right) =$ $v\left( \left|
u\right\rangle \left\langle u\right| ^{\left( 1\right) }\otimes {\bf 1}%
^{\left( 2\right) }\right) v\left( {\bf 1}^{\left( 1\right) }\otimes \left|
w\right\rangle \left\langle w\right| ^{\left( 2\right) }\right) =$ $v\left(
\left| u\right\rangle \left\langle u\right| ^{\left( 1\right) }\right)
v\left( \left| w\right\rangle \left\langle w\right| ^{\left( 2\right)
}\right) $, where the last equality reflects the fact that the proposition
represented by the projector $\left| u\right\rangle \left\langle u\right|
^{\left( 1\right) }\otimes {\bf 1}^{\left( 2\right) }$ corresponds to the
first particle being in the state $\left| u\right\rangle $, whatever the
state of the second particle.}; secondly, if in the singlet state we measure
the spin component of each particle along the same (arbitrary) direction,
the results are perfectly correlated (always opposite), and the same
relations must exist between the values of propositions in any deterministic
hidden variables theory, $v\left( \left( 1,0\right) ^{\left( 1\right)
}\right) =v\left( \left( 0,1\right) ^{\left( 2\right) }\right) $, $v\left(
\left( 1,1\right) ^{\left( 1\right) }\right) =v\left( \left( 1,-1\right)
^{\left( 2\right) }\right) $, etc... Therefore the left-hand side of (\ref
{e8}) can be written as follows: 
\begin{equation}
\label{e9}
\begin{array}{c}
v\left( \left( 1,0\right) ^{\left( 1\right) }\right) v\left( \left(
1,-1\right) ^{\left( 2\right) }\right) +v\left( \left( 0,1\right) ^{\left(
1\right) }\right) v\left( \left( 1,1\right) ^{\left( 2\right) }\right)  \\ 
+v\left( \left( 1,1\right) ^{\left( 1\right) }\right) v\left( \left(
1,0\right) ^{\left( 2\right) }\right) +v\left( \left( 1,-1\right) ^{\left(
1\right) }\right) v\left( \left( 0,1\right) ^{\left( 2\right) }\right)  \\ 
=\left[ v\left( \left( 1,0\right) ^{\left( 1\right) }\right) +v\left( \left(
0,1\right) ^{\left( 1\right) }\right) \right] \times \left[ v\left( \left(
1,1\right) ^{\left( 1\right) }\right) +v\left( \left( 1,-1\right) ^{\left(
1\right) }\right) \right] =1.
\end{array}
\end{equation}
The last equality in (\ref{e9}) is a consequence of condition (b) in the
2-dimensional space of the spin states of the first particle, {\it q.e.d.}

The first two vectors in eq. (\ref{e8}) are not orthogonal to the last two,
and the corresponding projectors do not commute; moreover, there is no
common eigenstate to the four projectors. Then, we can neither prepare a
system in a state free of dispersion for the four propositions, nor check
experimentally the relation by a simultaneous measurement of them; in this
sense this relation is different from those that implement condition (b) ($%
\sum\limits_{i=1}^4\left| {\bf u}_i\right\rangle \left\langle {\bf u}%
_i\right| ={\bf 1}\Rightarrow \sum\limits_{i=1}^4v\left( {\bf u}_i\right) =1$%
), for which there are states simultaneously free of dispersion for each
tetrad of compatible projectors, and which can (in principle) be
experimentally verified in any state. But in deterministic hidden variables
theories non-compatible observables can have well definite values in the
same individual system, and therefore eq. (\ref{e8}), although not
consequence of a resolution of the identity, is a legitimate relation
between hidden variables values, based on the properties of the singlet
state.

We now state our second no-go theorem:
\begin{quotation}
{\it There is no set of values }$v\left( {\bf u}_i\right) ${\it \ verifying
eqs. (\ref{e6}--\ref{e8}).}
\end{quotation}
The proof rests once more on a parity argument: each proposition appears
twice in (\ref{e6}--\ref{e8}), but the sum of the right-hand sides is 3.

We will finish with a reflexion on the number of propositions involved in
the proof; we have counted them explicitly using another consequence of
requisites (a) and (b):
\begin{enumerate}
\item[(d)]  {\it Given an individual system prepared in a state }${\bf w}$%
{\it \ and a set of vectors }$\left\{ {\bf u}_i\right\} $ {\it spanning a
subspace }$U${\it \ that contains }${\bf w}${\it , then }$%
\sum\limits_iv\left( {\bf u}_i\right) =1${\it .}
\end{enumerate}
\noindent A reason for this was given in \cite{6}: any vector ${\bf v}$ in
the subspace complementary to $U$ is orthogonal to ${\bf w}$; $v\left( {\bf w%
}\right) =1$ and (b) imply $v\left( {\bf v}\right) =0$; therefore the sum of
values of any complete set of compatible propositions on the subspace $U$
must be $1$. This can be justified too, without explicitly quoting the
values $v\left( {\bf v}\right) =0$, if we keep in mind that the probability
of finding the system in a state in the subspace $U$ is $1$. Condition (d)
is a consequence of (a) and (b), but in contradistinction to what happened
in (c), now the propositions omitted have definite zero values, and
therefore it could be argued that using (d) is essentially a way to count
the number of vectors appearing in state-specific proofs \cite{6}, leaving
out the vectors orthogonal to the initial state\footnote{%
Starting from a state-specific BKS\ proof in terms of $n$-dimensional
vectors we can trivially find a state-specific proof in any dimension $n+m$:
it suffices to append $m$ zeros to all vectors, including the initial state.
If the counting of the number of vectors is based on the explicit use of
condition (d), this number is independent of $m$ (for instance, the
10-vector set of the state-specific proof in dimension $n=4$ in ref. \cite{1}
gives also a 10-vector state-specific proof in $n=8$). The ratio between the
number $f$ of vectors used in a demonstration of the BKS theorem and the
dimensionality $n$ of the space \cite{6} is a reasonable measure of the
merit of a state-independent proof (where obviously $f>n$), but its meaning
in state-specific proofs is questionable if the initial state and the
vectors orthogonal to it are not counted, as it is done when using condition
(d).}.

In our example, both triads of vectors on the left-hand sides of (\ref{e6},%
\ref{e7}) span subspaces that contain the singlet $\left( 0,1,-1,0\right) $,
and therefore both equations are a direct application of rule (d). If we use
this condition, we only need to count the 5 vectors explicitly appearing in (%
\ref{e6},\ref{e7}); a count of the number of propositions with definite
values involved in the theorem, based strictly on (a) and (b), should also
include the omitted vectors $\left( 1,1,1,-1\right) $, $\left(
-1,1,1,1\right) $ (and perhaps the initial state too).

In the second part of this paper we have proved the impossibility of
assigning non-contextual values to a set of propositions in the singlet
state. The contradiction arises in only 3 equations involving 5 propositions
in ${\bf R}^4$: in terms of {\it number of propositions,} this
state-specific BKS proof is the most economic no-go theorem that we know of.\\

We would like to thank Rob Clifton for the private communication that got
this paper started, the anonymous referee for his comments on the second
part, and Asher Peres and Gabriel \'Alvarez for their patience and advice.

\end{document}